\documentclass[twocolumn,showpacs,preprintnumbers,superscriptaddress,nofootinbib]{revtex4}
\usepackage{graphicx}
\usepackage{dcolumn,amsmath,amssymb}
\usepackage{bm}\newcounter{n} \newtheorem{definition}{Definition}
\begin{document}


\title{INTRIGUING SYMMETRY IN STATISTICAL STRUCTURES OF SIBERIAN LARCH TRANSCRIPTOME}

\author{Michael G.\,Sadovsky}
\affiliation{Institute of computational modelling of SD of RAS;\\ 660036 Russia, Krasnoyarsk, Akademgorodok.} \email{msad@icm.krasn.ru}
\author{Vladislav V.\,Birukov}
\affiliation{Siberian Federal university;\\ 660041 Russia, Krasnoyarsk, Svobodny prosp.79.} \email{vladbir2010@gmail.com}
\author{Yuliya A.\,Putintseva}
\affiliation{Siberian Federal university;\\ 660041 Russia, Krasnoyarsk, Svobodny prosp.79.} 
\affiliation{Institute of forestry of SD RAS;\\ 660036 Russia, Krasnoyarsk, Akademgorodok.} 
\author{Konstantin V.\,Krutovsky}
\affiliation{Siberian Federal university;\\ 660041 Russia, Krasnoyarsk, Svobodny prosp.79.} \email{nvoreshkova@yandex.ru}
\affiliation{Institute of forestry of SD RAS;\\ 660036 Russia, Krasnoyarsk, Akademgorodok.} \email{nvoreshkova@yandex.ru}
\affiliation{Georg-August-University of G\"{o}ttingen, B\"{u}sgenweg, 2, G\"{o}ttingen, D-37077, Germany}
\affiliation{N.\,I.\,Vavilov Institute of General Genetics of RAS; Gubkin Str., 3, Moscow, 119333, Russia}
\affiliation{Texas A\,\&\,M University, HFSB 305, 2138 TAMU, College Station, Texas, 77843, USA}

\begin{abstract}
The paper presents a novel approach to infer a structuredness in a set of symbol sequences such as transcriptome nucleotide sequences. A distribution pattern of triplet frequencies in the Siberian larch (\textit{Larix sibirica}~Ledeb.) transcriptome sequences was investigated in the presented study. It was found that the larch transcriptome demonstrates a number of unexpected symmetries in the statistical and combinatorial properties.
\end{abstract}


\maketitle

\section{\label{introd}Introduction}
Key challenge of an up-to-date molecular biology, bioinformatics and mathematics is a search for an order and structuredness in bulk data provided by modern pipelines deciphering genomes. Genomics (and related fields of molecular biology) as well, as linguistics are the second to none in that respect. Indeed, amount of raw nucleotide sequence data grows daily for billions of megabytes. Those sequences are symbol sequences based mainly on the four-letter alphabet $\aleph=\{\mathsf{A}, \mathsf{C}, \mathsf{G}, \mathsf{T}\}$. A question on the order and structuredness in such biologically provided data depends rather heavily on the type and kind of the data. In general, the data could be referred as symbol sequences from four-letter alphabet; yet, one must check and validate the available data to proceed, before any comprehensive and substantial analysis of these former.

Here we studied an order and structuredness over a set of sequences representing the transcriptome of Siberian larch (\textit{Larix sibirica}~Ledeb.). Transcriptome represents sequences of expressed genes and corresponds to the \mbox{mRNAs} moleculae isolated from biological cells or tissues.

Two main (and complimentary, to some extent) approach may be persuaded to explore the structuredness and order in the data mentioned above. The former is based on the search of various regularities, unexpected sited, etc. within a sequence \cite{z1,z3}; this approach is of great importance for annotation of newly deciphered genomes (and not only). The latter stipulates each sequence in a set of entities under consideration to be an object (a ``point'', roughly speaking), and an order or structuredness here is understood as an inhomogeneity in the mutual location of those objects in some proper space, with correlation to inhomogeneities in the distribution of the objects observed over the other attributes \cite{z1-1,z2}. We follow this approach here.

Key idea in our search for a structure and order in a set of symbol sequences (a set of nucleotide sequences comprising larch transcriptome) is to transform sequences into their frequency dictionary \cite{bug96,bug97,bug98,kitaj}. There could be a number of various definitions of a frequency dictionary, but we will use the basic one that is a list of all the strings of a given length accompanied with a frequency of each string (a detailed description is given below). It is crucial that the transformation of a symbol sequence into a frequency dictionary allows us to map a set of sequences into a metric space. The latter provided us with powerful and extended tools for analysis. For our further analysis we also assumed that neither other symbols, nor blank spaces are supposed to be found in a sequence; a sequence under consideration is also supposed to be coherent (i.\,e. consisting of a single piece).

Usually, a distance between sequences is used to find out regularities or similarities among them. Distances between sequences are most often based on sequence alignments (see \cite{similarities} for the first computer algorithm for aligning two sequences and \cite{alignment} for a recent review). However, there could be serious problems and constraints in generating reasonable alignments for highly divergent sequences. Meanwhile, there are other concepts and methods that could be much more powerful than those that are based on alignments (e.\,g., \cite{znam1,znam2}), although they still need further investigation of their applicability for addressing biological problems.

We will briefly outline the concept of our study and then demonstrate the main results obtained. First, we changed each symbol sequence (that is a nucleotide sequence in the Siberian larch transcriptome set) into a frequency dictionary. Then, we studied distribution of those dictionaries in a multidimensional metric space trying to infer any inhomogeneities, regularities and clusters in this distribution. This clustering was carried out using the $K$-means technique. As we have found, a clusterization is rather clear, distinct and stable.

Second, we compared the statistical properties of the clusters identified by $K$-means and found that these clusters demonstrated a very strong symmetry in terms of the statistical properties. In brief, being checked against each other, the clusters showed extremely low level of discrepancy in the Chargaff's second parity rule. This low discrepancy is the most intriguing fact concerning the properties of the studied transcriptome sequence set. On the contrary, the discrepancy determined within these two classes separately has been found to be rather high. Such significant loss of Chargaff's parity discrepancy may follow from the occurrence of the sequences representing the transcriptome at the opposite strands, and this idea has been approved with BLAST.
\begin{figure}
\includegraphics[width=8.2cm]{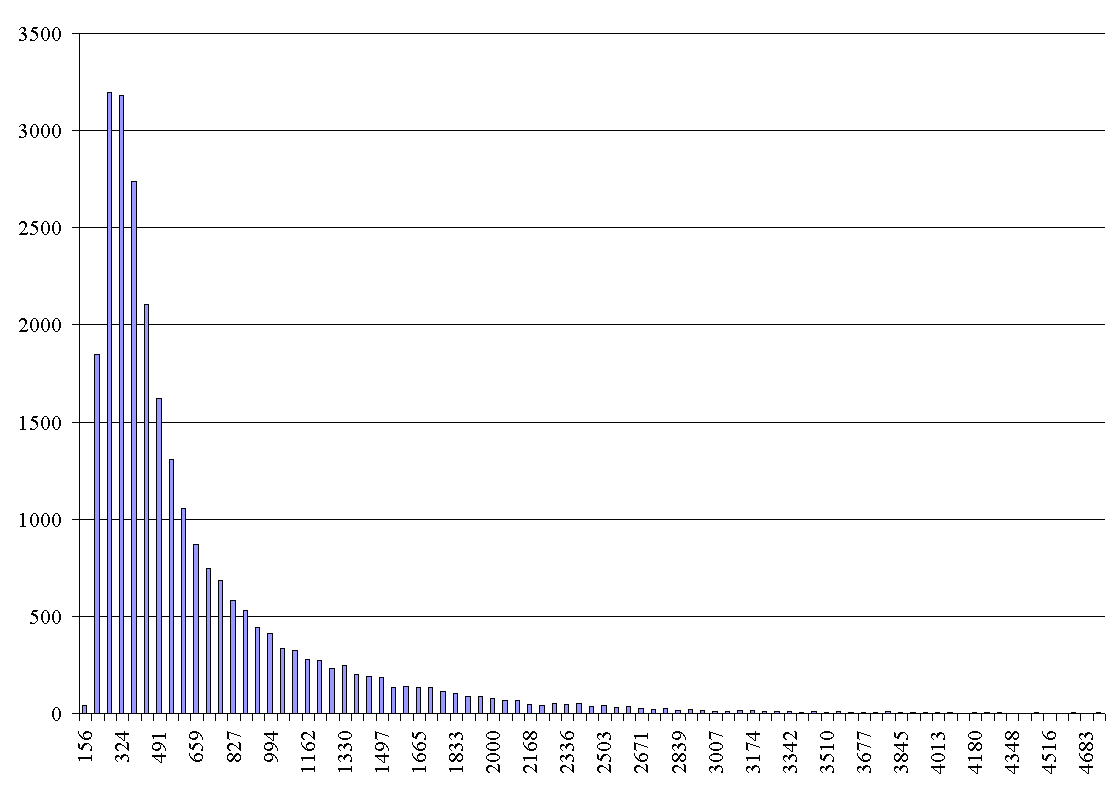}
\caption{\label{fig1}Histogram distribution of the transcriptome sequences over their length. The case of all 43\,686 entries.}
\end{figure}

Finally, it has been found the transcriptome to yield an unusual octahedral pattern, when the distribution of the specially implemented frequency dictionaries was developed. Unlike the originally found seven-cluster structure \cite{z1,z3}, here the distribution of the sequences arranged an octagon, with six distinctively identified vertices.

\section{\label{matmet}Materials and Methods}
\subsection{Transcriptome nucleotide sequence data}\label{seqdata}
\begin{figure}
\includegraphics[width=8.2cm]{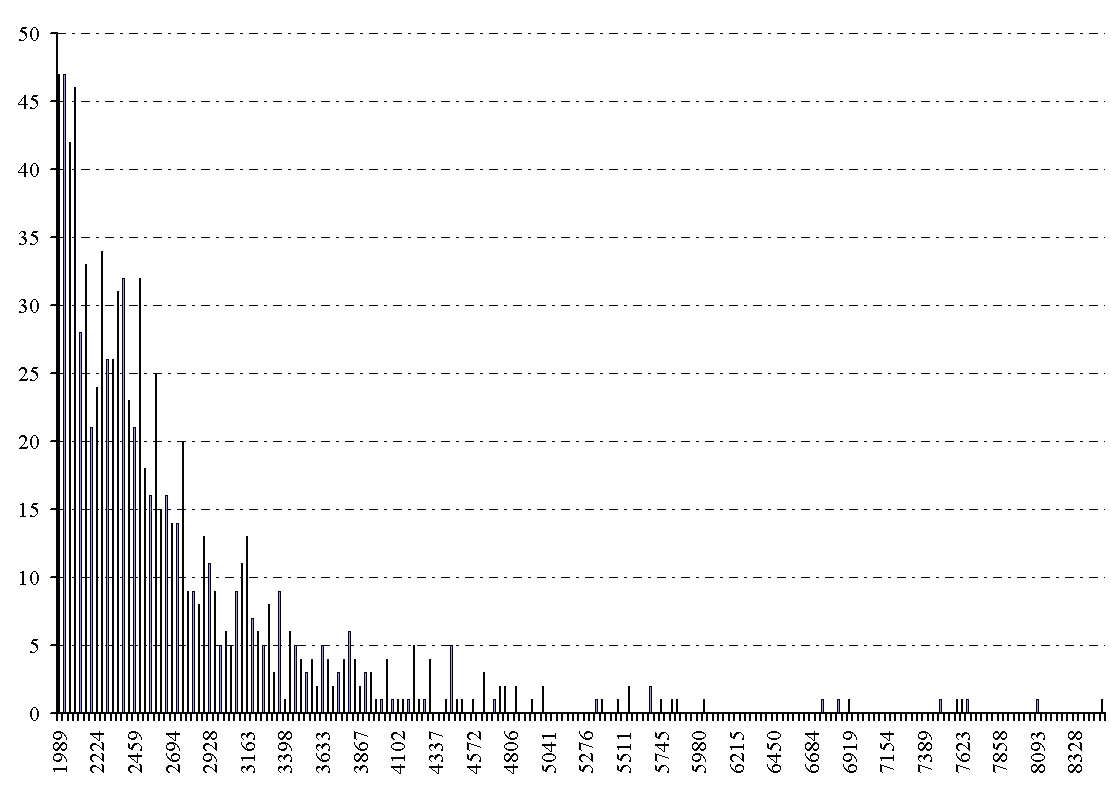}
\caption{\label{fig2}Histogram distribution of the transcriptome sequences longer $2\times 10^3$ symbols over their length; totally, 1436 entries.}
\end{figure}

The transcriptome of Siberian larch was originally sequenced in the project on the whole genome sequencing of Siberian larch \cite{krutovsky1}. The sequence data of \textit{L. sibirica} were obtained using Illumina MiSeq sequencer at the Laboratory of forest genomics of the Siberian Federal University. Total RNA was isolated from buds \cite{krutovsky2}. Total number of sequences in the transcriptome set was 43\,686. The shortest sequence had 201 nucleotide base pairs (symbols), while the longest one had 8\,512~symbols. An average length of the sequences in the transcriptome was $606.44$~bp, with the standard deviation of $609.28$~bp. Here we cut off the strings shorter than 200~b.\,p., for the analysis. Of course, it should be said that the sequencer facility yields shorter strings, as well. The histograms of the distribution of the transcriptome sequences entries over their length are presented in Fig.~\ref{fig1}. Evidently, the distribution resembles quite strongly Poisson distribution. We studied the sequences longer $2\times 10^3$ symbols. Fig.~\ref{fig2} shows the distribution of the sequences longer $2\times 10^3$ symbols over their lengths.

\subsection{Frequency Dictionary}
Previously \cite{bug96,bug97,bug98,kitaj}, a frequency dictionary was proposed to be a fundamental structure of a symbol sequence. Here we introduce the definition of frequency dictionary. First, consider a symbol sequence $\mathfrak{T}$ of the length $N$ from the four-letter alphabet. Hereafter we assume that no other symbols but those from the alphabet $\aleph=\{\mathsf{A}, \mathsf{C}, \mathsf{G}, \mathsf{T}\}$ are found in a text~$\mathfrak{T}$, and there are also no gaps in the text. The following are a few definitions:
\begin{definition}
The \emph{\textbf{word}} $\omega_q = \nu_1\nu_2\ldots\nu_{q-1}\nu_q$ of the length~$q$ is a string occurred in the text~$\mathfrak{T}$. Here~$\nu_j$ is a symbol occupying the $j$-th position at the word; $\nu_j \in \aleph$.
\end{definition}
Any word $\omega_q$ is identified by the window of the size~$q$. Any position of such window found alongside a text~$\mathfrak{T}$ determines a word. Let \textbf{step}~$t$ be the distance between two neighbouring positions of the window; obviously, $1\leqslant t < N$ (here $N$ is the length of $\mathfrak{T}$.
\begin{definition}\label{freqdic}
Frequency dictionary $W_{q,t}$ is the set of all the words of the length~$q$ counted within the text~$\mathfrak{T}$ with the step~$t$ so that each word is accompanied with its frequency~$f_{\omega}$. A frequency of a word~$\omega$ is defined traditionally: that is the number~$n_{\omega}$ of copies of the word~$\omega$ divided the total number~$M$ of all copies of all the words.
\end{definition}
As a rule, frequency dictionary is understood as $W_{q,1}$ dictionary \cite{bug96,bug97,bug98,kitaj}. Everywhere below we shall consider the dictionaries~$W_{3,1}$ and~$W_{3,3}$.

Frequency dictionary~$W_{3,1}$ ($W_{3,3}$ either) unambiguously maps a text~$\mathfrak{T}$ into a 64-dimensional space, where the triplets are coordinate axes in those space, and the frequencies are the coordinates. Hence, frequency dictionary represents a short range (or meso-scale, at most) structuredness in a symbol sequence. That is the basic issue for further analysis of statistical properties of a symbol sequence representing the \textit{L. sibirica} transcriptome. The key idea of the study was to check whether the sequences corresponding to various mRNA moleculae differ in their statistical properties or not.

Mapping of a sequence provided by frequency dictionary converts a sequence into a point in metric space. It means, that a distance between two points is defined. Everywhere further, we shall use standard Euclidean distance:
\begin{equation}\label{eqdist}
\rho\left(W_3^{(1)}, W_3^{(2)}\right) = \sqrt{\sum_{\omega = \mathsf{AAA}}^{\mathsf{TTT}}\left(f_{\omega}^{(1)}-f_{\omega}^{(2)}\right)^2}\,.
\end{equation}
The index $\omega$ enlists the triplets at the frequency dictionary.

Frequency dictionary $W_{3,t}$, indeed, maps a sequence into 64-dimensional metric space. In fact, that is not so, strictly speaking. The point is that the sum of all frequencies
\begin{equation}\label{summa_ch}
\sum_{\omega = \mathsf{AAA}}^{\mathsf{TTT}} f_{\omega} = 1\,.
\end{equation}
Here index $i$ enlists the triplets (that are $64 = 4^3$ in number). Such linear constraint makes the frequencies of triplets dependent; an implementation of any method for clusterization (visualization, etc.) may be deteriorated with the constraint. A linear constraint~\eqref{summa_ch} may cause an appearance of some additional, parasitic signal which may have nothing to do with the real clusterization.

Actually, the points corresponding to sequences under consideration are located in the linear space of co-dimension~1. Thus, a real distribution of the points is arranged in 63-dimensional space, and a triplet must be excluded from the analysis. Formally speaking, any of 64 triplets may be excluded: the constraint~\eqref{summa_ch} makes no preferences in such triplet choice.

Practically, a choice of the triplet to be excluded is not free. Assume, there is a triplet yielding absolutely equal frequency, over the entire set of genomes enlisted in the database. Then, this triplet makes no contribution into the discretion of the objects. This fact allows to formulate the rule to choose a triplet to be excluded. The rule sounds as follows: \textit{the triplet yielding the least standard deviation determined over the database must be excluded}. This rule means that the triplet with the least contribution to the discretion of objects is to be omitted. We used this one rule, in our studies. Meanwhile, one can figure out few more rules for the exclusion, and they do not contradict each other. We shall not discuss here any more a competition of such rules.

\subsection{Clustering techniques}\label{slust-met}
We used $K$-means technique to analyze the transcriptome. $K$-means technique is a good method for analyzing data of various nature (see \cite{fukunaga}). However, the following definitions should be additionally presented beforehand:
\begin{definition}\label{center-def}
Center of a class developed due to $K$-means is the arithmetic mean of the numbers describing the objects, to be calculated within a class.
\end{definition}
Of course, both a $K$-means implementation and the definition of a center depend on the metrics used to do it. We used Euclidean distance as a metric in our study.
\begin{definition}\label{radius-def}
Radius of a class is the arithmetic mean of the distances calculated from each point of a class against its center.
\end{definition}
High popularity of $K$-means yet does not make it free from a few following problems:
\begin{list}{\arabic{n})}{\usecounter{n}\leftmargin=9mm \labelwidth=5mm \topsep=0mm \labelsep=2mm \itemsep=1pt \parsep=0mm \itemindent=-1pt}
\item stability of a final distribution;
\item the number of classes determination, and
\item separability of classes.
\end{list}
We discuss them here in more detail. Since implementation of a clustering by $K$-means starts from a random dispersion of the original data set for $K$ classes, then one may not be sure that the final composition of the classes would remain the same in a new run of clustering. Of course, one might face the situation when the final distribution is identical for any initial separation. This is the situation of the highest stability of clustering. Looking ahead, we have found a strong stability of clustering done over the transcriptome.

On the contrary, there might be a situation where any new run of the procedure brings absolutely another final distribution. This is an instability case, and such instability could be a signature of a total lack of any intrinsic structuredness in the data set. In reality, the situation could be somewhere in between. As a rule, a set of data is divided into two subsets: the former tends to yield rather stable distribution in a series of $K$-means runs, and the latter gathers the objects that always change their class attribution. These are so called volatile objects. There is no simple or evident way to deal with them. An elimination of them from the original data set may cause a loss of the stability of a new classification for the rest of a data set. Here one may assume a failure of the method; otherwise, they should be considered as a separate set to be specially studied.

An implementation of a classification through $K$-means described above is not a final step yet. One must check whether the obtained classes are distant enough. This problem is tightly related to the problem of a number of classes to be developed by $K$-means. Indeed, there is no a priori way to figure out the exact number of classes for $K$-means classification; it is a matter of expertise of a researcher, as a rule. Thus, an advanced technique protocol implies that one starts from a sufficiently large number of classes, and then the number is decreasing through the amalgamation of indistinguishable classes.
\begin{figure}
\includegraphics[width=8.2cm]{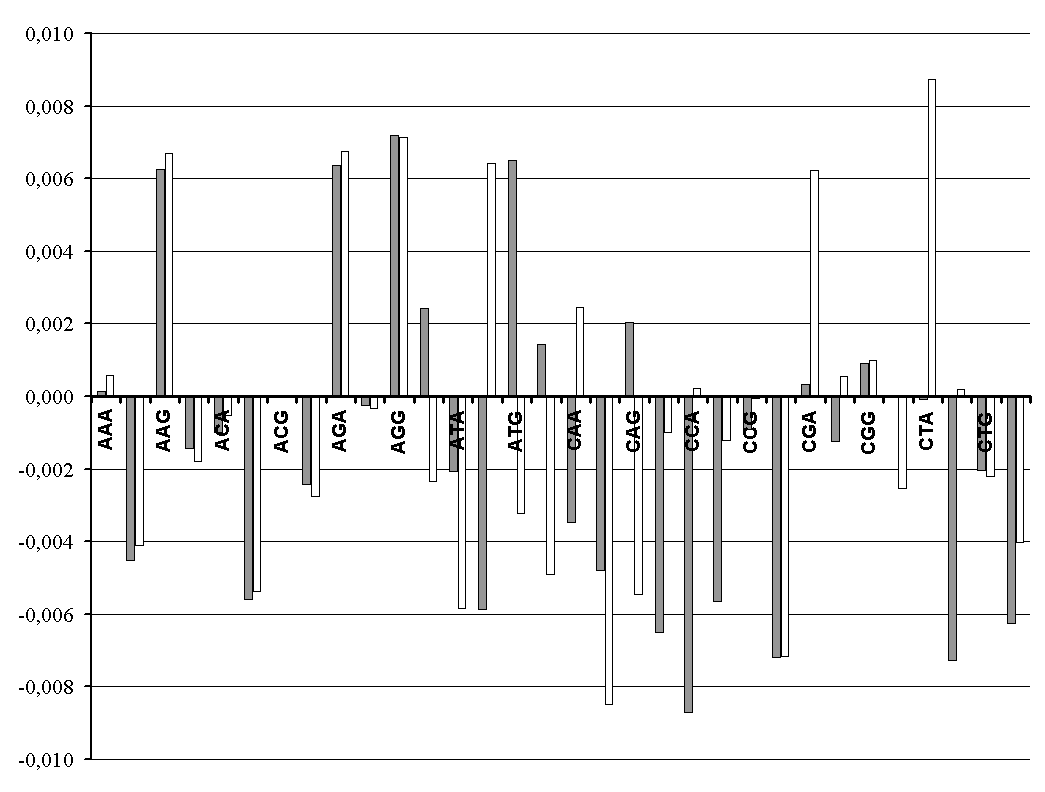}
\caption{\label{fig3}Discrepancy $\mu$ \eqref{chrg1} determined for center within two classes. Here 32 couples are present.}
\end{figure}

There are various criteria to distinguish classes \cite{book}. The strongest one assumes that two classes are discernible, if a distance between the classes is not less than the sum of their two radii. On the contrary, the weakest one requires that the greater radius does not exceed the distance between the classes.\label{page_criter} If two classes are indiscernible, then they must be amalgamated, and $K$-means must be run again, with a new number of classes decreased by one. Thus, the advanced version of the method does not increase the number of classes and stops at the maximal set of distinguishable classes. Here we will demonstrate that the classes observed over the transcriptome in this study were highly distinguishable.

\subsection{Elastic map technique}
This technique aims, first of all, to visualize multi-dimensional data. The approach is based on the approximation of multi-dimensional data (of high dimension) with a manifold of low dimension; in particular, we used two-dimensional manifold for the approximation. In brief, the method consists of the following steps. At the beginning \cite{z1-1}, two principal components must be determined. Then a plane must be developed over these two principal components, as on the axes. At the next step, each data point must be projected on the plane, and the projection should be connected with the original point with an elastic spring.
\begin{figure}[!t]
\includegraphics[width=8.4cm]{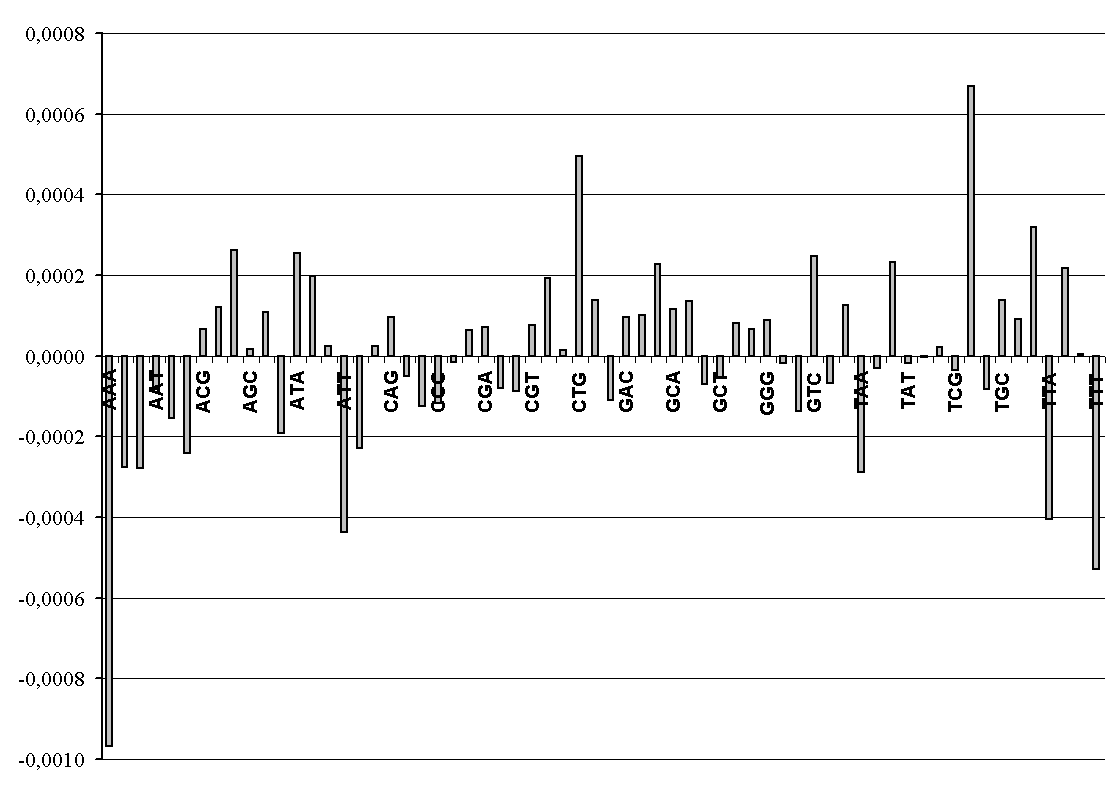}
\caption{\label{fig4}Discrepancy $\mu$ \eqref{chrg1} determined between the centers of two classes. Here 64 couples are present.}
\end{figure}

As soon, as this construction is ready, one should change the plane for elastic surface, that can bend, expand and shrink. The obtained construction is released to reach the minimum of the potential energy: initially tightened springs shrink off, but the surface is expanding and bending, thus consuming energy. As the construction reaches the configuration corresponding to the energy minimum, the data points must be redefined, on the jammed surface.

\begin{figure*}
\includegraphics[width=17.2cm]{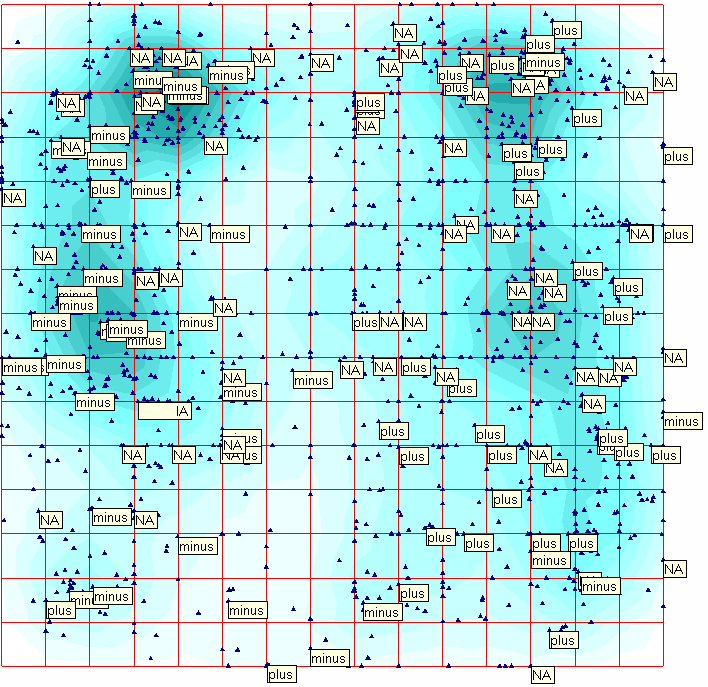}
\caption{\label{fig5}Distribution of sequences over the elastic map. Labels \textsl{plus} and \textsl{minus} indicate the sequences belonging to leading (\textsl{plus}) and lagging (\textsl{minus}, respectively) strands determined by BLAST.}
\end{figure*}

This redefinition consists in the following. Each data point is now mapped into the point on the surface that is the closest one (in terms of the metrics used in the study), instead of the orthogonal projection used initially. Finally, the jammed surface (elastic map) is getting smoothed out to look like a plane. Here is the end.

On the contrary to $K$-means, here one may not expect to develop a series of classifications consisting of different number of classes; elastic map technique provides a unique clusterization of the multi-dimensional data. Few words should be said towards the local density colouring. This is very useful and informative tool to analyze the data. To develop the colouring, one must supply each data point image at the surface with the Gaussian function centered at the point image; a specific width (equivalent to standard deviation of a normal distribution function) of the function is a free parameter chosen by researcher.

As soon, as each point at the map is supplied with the function, they are to be summed up; the result function is presented as colored layers.

Again, the clusterization provided by elastic map technique is not free from an arbitrariness brought by a researcher: there is the key parameter that can not be determined automatically; a specific semi-width of the Gaussian function covering each point is that latter. The specific value for this parameter is to be put on by a researcher; of course, software provides a by-default value, meanwhile that is not the best choice, in any case. Everywhere in our studies we chose the parameter equal to $0{,}15$ instead of $0{,}25$ by-default value.

\section{Results}\label{rez}
\subsection{Classification with $K$-means}
We developed consequently four classifications using $K$-means technique. The number of classes varied from two to five. For each number of classes 350 runs of $K$-means were executed to study the stability of classification. The stable subsets of sequences comprising the transcriptome were determined for each classification. We assumed a classification to be stable, if not less than 95\,\% of all runs yielded the same distribution of sequences. Moreover, the dictionaries comprising the stable core of the classes are not the only character of a classification; another issue is the behaviour of the volatile dictionaries. These latter may follow two opposing patterns: either to change a class attribution ``all together'', or change it almost randomly. The first option was realized, in our case. All four classifications were very stable for 350~runs.

Moreover, the radii of classes were calculated for each classification, as well as the distances between them (see Definitions~\ref{center-def} and~\ref{radius-def} above).

\subsection{Chargaff's parity rule for two classes}
Classification of the sequences of \textit{L. sibirica} transcriptome (longer $2\times 10^3$ symbols) by $K$-means is of great importance; let concentrate on it in more detail. The triplet $\mathsf{GAC}$ has been eliminated, for classification implementation. To begin with, the classification was very stable (see paragraph above); besides, it yields a good separability of classes (see Section~\ref{razlichimost} below). Chargaff's (generalized) second parity rule stipulates a proximity of frequencies of two words comprising a complimentary palindrome \cite{chargaf1,chargaf2}.
\begin{figure}[!t]
\includegraphics[width=8.4cm]{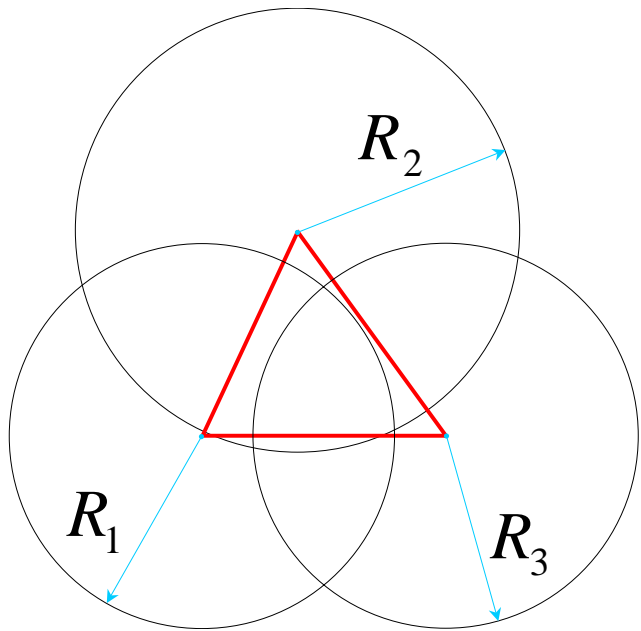}
\caption{\label{fig_rad3}Distance between the classes for $K$-means clusterization; $K=3$. $R_1 = 0{,}027184$, $R_2 = 0{,}031474$, $R_1 = 0{,}027140$; $\rho\left(C_1, C_2\right)= 0{,}031960$, $\rho\left(C_1, C_3\right)= 0{,}035818$, $\rho\left(C_2, C_3\right)= 0{,}034378$.}
\end{figure}
\begin{definition}
Couple of words (of the length~$q$ that are read equally in opposite directions, with respect to the substitution rule (prescribed by the first Chargaff's rule) makes \emph{\textbf{complimentary palindrome}}.
\end{definition}
For example, the couples $\mathsf{AAA}\Leftrightarrow \mathsf{TTT}$, $\mathsf{ACG}\Leftrightarrow \mathsf{CGT}$, \[\mathsf{CGAATACG}\Leftrightarrow \mathsf{CGTATTCG}\] are the complimentary palindromes. In any frequency dictionary of a thickness~$q$ there always exists $\frac{1}{2}4^q$ couples of complimentary palindromes; in particular, for triplet frequency dictionaries there always exist 32 couples of such triplets. Not discussing here the origin and properties of nucleotide sequences manifesting in the second Chargaff's parity rule, we just mark up the fact that various genetic entities possess specific figures of the discrepancy that characterizes the entity, from the point of view of the exactness of a feasibility of that former (see details in \cite{chargaf1}). Possible way to define the discrepancy~$\mu$ is following \cite{chargaf1}:
\begin{equation}\label{chrg1}
\mu = \dfrac{1}{|\Omega|}\sqrt{\sum_{\omega \in \Omega}\left(f_{\omega} - f_{\overline{\omega}}\right)^2}\,.
\end{equation}
Here $|\Omega|$ means a set of complementary palindromes, while $\omega$ and $\overline{\omega}$ are the words making the complementary palindrome; $|\cdot |$ means the capacity of a set. The figure~$\mu$ defined according to~\eqref{chrg1} looks like a distance, meanwhile it is not. It does not provide a measure between two points, but presents a discrepancy of a given dictionary.

Chargaff's second parity rule might look rather mysterious, while it does not. Indeed, suppose the substitution rule (similar to Chargaff's first rule) holds true, for some sequences from a proper alphabet (which must be even, at least). Let here concentrate on the four-letter alphabet $\aleph$; so the substitution rule looks like $\mathsf{A}\Leftrightarrow \mathsf{T}$, $\mathsf{C}\Leftrightarrow \mathsf{G}$. Suppose then a sequence from the alphabet is long enough, and random\footnote{Here we shall not discuss what is a random finite sequence, while relevant strict definitions and essential concepts are well known.}, and the probabilities ($=$ frequencies) are as following:
\begin{equation}\label{bernul}
p(\mathsf{A})=p(\mathsf{T})= \alpha;\quad p(\mathsf{C})=p(\mathsf{G})= 1-\alpha\,.
\end{equation}
Then, due to a randomness of the sequence, the frequencies of two strings comprising a complimentary palindrome are exactly the same.

Evidently, a real genetic sequence is rather far from any random one; of course, due to a finiteness of that latter, one always can exactly and unambiguously match a (finite) sequence with Markov process realization of a proper order \cite{bug96,bug97,bug98,s1}, but we shall not discuss this point any more here. The fact is that any real DNA sequence exhibits a lot of deviations from a random sequence. Thus, a question arises what is a figure of violation of the measure~\eqref{chrg1}, observed for random sequences, if the first rule constraint~\eqref{bernul} is violated with some given level $\lambda \ll 1$. It was found the discrepancy $\mu$ to remain the same, with some constant $C$, for any longer words, as $q>1$ \cite{chargaf1}: $\mu_q \sim C\cdot \lambda$. It has been found also the discrepancy~\eqref{chrg1} goes down, as $q$ grow up, for real sequences.

Thus, we have checked the pattern of the second Chargaff's parity rule feasibility observed over the transcriptome contigs ensemble, for three cases:
\begin{list}{--}{\leftmargin=6mm \labelwidth=5mm \topsep=0mm \labelsep=2mm \itemsep=1pt \parsep=0mm \itemindent=-1pt}
\item within the first class identified through $K$-means;
\item within the second class identified through $K$-means, and
\item between these two classes.
\end{list}
\begin{figure}
\includegraphics[width=8.4cm]{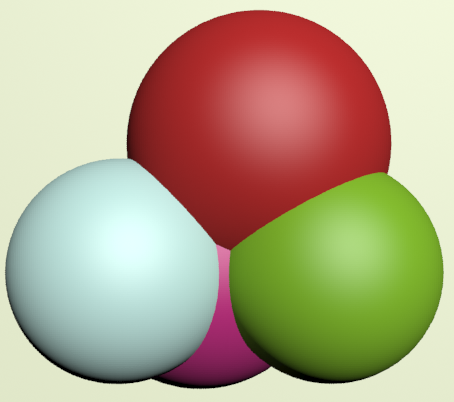}
\caption{\label{4klssa_1}Distribution of the classes, for $K=4$; the figure illustrates Table~\ref{tab2}.}
\end{figure}
Everywhere here the discrepancy~\eqref{chrg1} was calculated for the centers of classes, not for individual sequences. The most surprising thing was that the second (generalized) Chargaff's parity rule had significantly less discrepancy $\mu$ figure in the third case (the comparison of the centers of two classes of the sequences comprising the transcriptome). Figures \ref{fig3} and \ref{fig4} illustrate this fact.

\subsection{Classes discernibility}\label{razlichimost}
That is a common place that the implementation of $K$-means with $K>2$ classes is not the end, yet. Indeed, one should check the discernibility of the classes obtained with the classification. A discernibility may be defined with strong constraint for the class discernibility, and with the weak one (see Subsection~\ref{slust-met}, page~\pageref{page_criter}, specifically, for the criteria of discernibility).
\begin{figure}
\includegraphics[width=8.4cm]{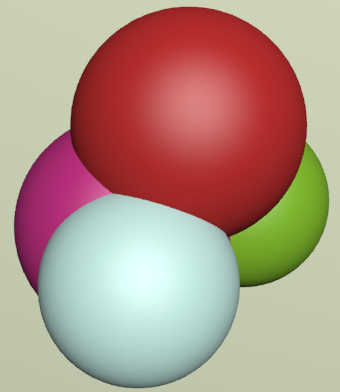}
\caption{\label{4klssa_1}Distribution of the classes, for $K=4$; the figure illustrates Table~\ref{tab2} (another aspect).}
\end{figure}

We have checked all the classes obtained with $K$-means ($2 \leqslant K \leqslant 8$) for the discernibility. Namely, the distance (pairwise) between the class centres and the class radii have been calculated. Considerable level of discernibility has been observed for two-class distribution: the distance between the classes is equal to $\rho\left(C_1, C_2\right)= 0{,}035145$, with the corresponding radii equal to $R_{C_1} = 0{,}026931$ and $R_{C_2} = 0{,}027204$; hence, an excess of the sum of radii over the distance~$\rho$ is equal to $0{,}018990$. The distances between the centres of the classes and the radii, for the cases of $K=3$ are shown in Fig.~\ref{fig_rad3}.

Similar figures for four-class and five-class distributions are shown in Table~\ref{tab2} and Table~\ref{tab1}, respectively. As usual, a growth of class number results in a growth of class discernibility: for $K=5$ class one exhibits an excess of a distance between all the classes over the sum of the relevant radii, except the class two (see Table~\ref{tab1}).
\begin{table}
\begin{tabular}{r|c|c|c|c}\hline
$C_j$ & $R_{C_{j}}$ & $\rho\left(C_{j}, C_2\right)$ & $\rho\left(C_{j}, C_3\right)$ & $\rho\left(C_{j}, C_4\right)$\\\hline
1 & $0{,}025631$ & $0{,}047767$ & $0{,}026092$ & $0{,}044014$\\
2 & $0{,}026911$ &  & $0{,}035593$ & $0{,}034374$\\
3 & $0{,}024011$ &  &  & $0{,}034092$\\\hline
\end{tabular}
\caption{\label{tab2}The distribution of distances between the centres of four classes. $R_{C_4} = 0{,}031391$.}
\end{table}

\section{Discussion}\label{disk}
Surprisingly high symmetry in statistical structure of the \textit{L. sibirica} transcriptome has been found. Obviously, it could be the only reason behind such symmetry: a lot of the pairs of strings of ``sense-antisense'' type among the sequences in the set. The question is what is a source of those pairs of sequences. Fig.~\ref{fig5} answers the question.
\begin{table}[!h]
\begin{tabular}{r|c|c|c|c|c}\hline
$C_j$ & $R_{C_{j}}$ & $\rho\left(C_{j}, C_2\right)$ & $\rho\left(C_{j}, C_3\right)$ & $\rho\left(C_{j}, C_4\right)$ & $\rho\left(C_{j}, C_5\right)$\\\hline
1 & $0{,}025554$ & $0{,}034471$ & $0{,}055149$ & $0{,}056565$ & $0{,}055615$ \\
2 & $0{,}022914$ &  & $0{,}042883$ & $0{,}048523$ & $0{,}038412$ \\
3 & $0{,}025526$ &  &  & $0{,}051196$ & $0{,}039320$ \\
4 & $0{,}029491$ & &  &  & $0{,}042635$ \\\hline
\end{tabular}
\caption{\label{tab1}The distribution of distances between the centres of five classes. $R_{C_5} = 0{,}027268$.}
\end{table}

Genomes of organisms of various taxonomy rank differ significantly in the discrepancy of Chargaff's generalized second parity rule based on formula~\eqref{chrg1}. Mitochondria takes the leading position in terms of a level of violation of the second generalized parity rule. However, in general, there is less discrepancy of the rule for genome at a higher taxonomy rank \cite{chargaf1}. Chloroplasts are, in general, the second to mitochondria.

\begin{figure}[!t]
\includegraphics[width=8.2cm]{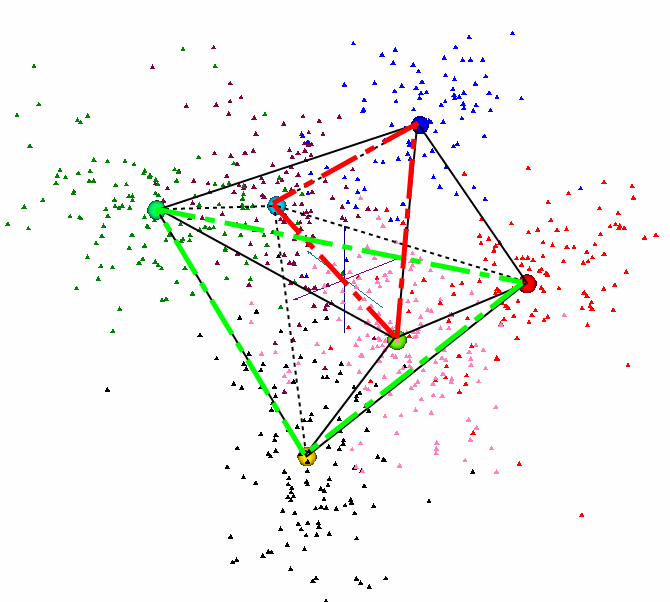}
\caption{\label{fig6}Octahedral distribution of contigs (frequency dictionaries, indeed), in 63-dimensional space of triplet frequencies; $\mathsf{CGC}$ triplet has been eliminated. The distribution is shown in principal components coordinates.}
\end{figure}

It is unlikely that the complementary pairs would result somehow in the process of cDNA synthesis due to some technical reasons. Alternatively, such composition of the set may arise from expression of genes in opposite directions in the complementary strands of the original DNA. This hypothesis would require genes in the \textit{L. sibirica} genome to have opposite orientation in both genomic DNA strands. Such genome arrangement is known for many (if not most) organisms, but it is mostly asymmetrical (e.\,g., \cite{trikitajza,chargaf2}). However, the key question here is the ratio of the genes in opposite directions in the complementary strands in \textit{L. sibirica}. A general estimation of the ratio of the opposite genes varies from $10\div 1$ to $1\div 1$, for various organisms and different authors.

It should be also stressed out that the very low discrepancy based on formula~\eqref{chrg1} observed for the centers of two classes mentioned above requires that a number of opposite genes located in complementary strands must also overlap. Such condition is unlikely, but it does not mean that such overlapping is not possible. Evidently, the first step to verify it may consist in checking the sequences belonging to various classes (obtained due to $K$-means technique) by BLAST. This can yield a list of sequences homological to the main strand, and to the complementary one, correspondingly. Such verification may also bring another advantage: it is a common fact that BLAST is a very time-consuming procedure. A combination of $K$-means technique to figure out the tentative specific strand strings with BLAST may seriously decrease the resource demand due to the specific pre-treatment of a set to be checked. However, these issues are beyond the scope of this paper and require additional studies.

\subsection{Octahedral structure of the transcriptome}
Previously \cite{z1,z3} basic 7-cluster structure to be observed in nucleotide sequences has been reported. Surprisingly, similar structure has been found in the transcriptome. To do that with the transcriptome, we are to consider the frequency dictionary $W_{q,t}$ with $t>1$ instead of the dictionary $W_{q,1}$ (see Definition~\ref{freqdic}). Indeed, the dictionary $W_{3,3}$ is under consideration, to reveal the clusterization described in~\cite{z1,z3}.

Unlike in the papers~\cite{z1,z3}, here we did not make an ensemble of fragments of (the same) sequence. Originally, the 7-cluster structure in bacterial (and yeast) genomes has been found through the comparative study of the frequency dictionaries of the overlapping fragments of some specific length~$L$ selected within a genetic sequence. Thus, each fragment (of a typical length close to 300~b.\,p.) yields a point in 63-dimensional space. Further clusterization of these points reveals a pattern with seven nodes: three of them correspond to the fragments falling inside the coding regions, other three ones correspond to the reciprocal fragments (roughly saying, to those to be found in the opposite strand), and the seventh one corresponds to the non-coding regions.

We did not identify any fragments within a sequence from the transcriptome ensemble; on the contrary, we have developed the same number of frequency dictionaries, while right now these are $W_{3,3}$ dictionaries, instead of $W_{3,1}$ ones. In other words, each transcriptome sequence (those under consideration, i.\,e. longer $2\times 10^3$ symbols, to be exact) is considered as a fragment as described in~\cite{z1,z3}. Fig.~\ref{fig6} shows the results of the clusterization.

The most important difference observed over the transcriptome is that the pattern found due to clusterization is the octahedron, not any other figure as described in~\cite{z1,z3}. Considering the contigs comprising the transcriptome (the longer subfamily of that latter, to be exact) as the fragments, one has to label them with respect to their ``relative phase''\!.

The relative phase determines a location of the start codon, within a sequence. Obviously, one can develop three different frequency dictionaries $W_{3,3}$ differing in the location of the very first triplet; this difference is usually called \textsl{reading frame shift}. Since the triplets in $W_{3,3}$ do not overlap, then there are three different tilings referred with the position of the very first nucleotide at the sequence:
\begin{list}{\arabic{n})}{\usecounter{n}\leftmargin=9mm \labelwidth=5mm \topsep=0mm \labelsep=2mm \itemsep=1pt \parsep=0mm \itemindent=-1pt}
\item starting immediately at the first symbol of the sequence ($W_{3,3}^{(0)}$);
\item starting at the second symbol of the sequence ($W_{3,3}^{(1)}$), and
\item starting at the third position ($W_{3,3}^{(2)}$)\,.
\end{list}
Here the upper index enumerates the positions of the very first triplet to be counted.

That is a common fact that any genetic sequence consists of the coding and non-coding regions. To simplify explanation, further we shall stipulate that such regions do not overlap. Coding and non-coding regions differ from the point of view their statistical properties: a set of three frequency dictionaries
\begin{equation}\label{3clov}
\left\{W_{3,3}^{(0)}, W_{3,3}^{(1)}, W_{3,3}^{(2)}\right\}
\end{equation}
determined for coding region differs strongly from a similar set derived over a non-coding region. Further, we shall not consider all three possible dictionaries~\eqref{3clov}; instead, we shall consider the frequency dictionary $W_{3,3}^{(0)}$ always.
\begin{figure}[!t]
\includegraphics[width=8.2cm]{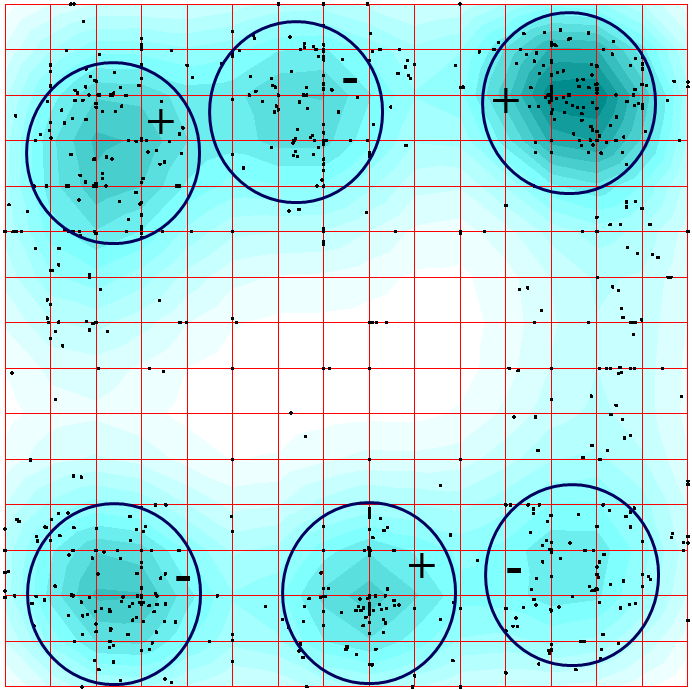}
\caption{\label{fig7}Six-cluster distribution of the contigs; those from leading strand are labeled with \textsl{plus} sign, while those from the lagging one are signed with \textsl{minus} sign.}
\end{figure}

Since the development of $W_{3,3}^{(0)}$ dictionary always takes start from the very first symbol of a sequence, then one has to learn whether this first symbol matches exactly the border of a coding region (or a gene, or another genetic entity which is frame shift sensitive). Yet, there is no way to do it in advance. We may assign each sequence from the ensemble with so called \textsl{relative phase} figure. That latter is equal to~0 to those sequences which take start exactly at the border of a coding region. The sequences that are shifted for a symbol upright, from the border, have the phase figure~1, and finally, the sequences with relative phase equal to~2 take start from the third symbol, within a coding region.

Yet, we have no way to identify the sequences comprising the transcriptome ensemble with the exact location alongside the chromosomes, with respect to borders of coding regions. The only point is that one should expect that the sequences possessing the same figure of relative phase would comprise a cluster, in triplet frequency space; Fig.~\ref{fig6} illustrates such distribution. Basically, there is no the seventh cluster, at all, in this pattern. An examination of the distribution over an elastic map (see Fig.~\ref{fig7}) also reveals that fact: there are six distinct clusters on the map.

Signs \textsl{plus} and \textsl{minus} indicate the clusters gathering the sequences from leading strand (\textsl{plus}) and those from lagging one (\textsl{minus}). This attribution to leading and lagging strands, as well as the relative phase impact is also proven with the similar clusterization provided over the set of $W_{3,3}^{(1)}$ and $W_{3,3}^{(2)}$ frequency dictionaries; yet, it brings no new pattern but the rotational permutation of the clusters on the map.

Indirectly, Fig.~\ref{fig7} proves a high quality of transcriptome sequencing and assembling: the fragments of the original sequence corresponding to the non-coding areas usually comprise the seventh cluster located elsewhere, but discretionarily. Here we see no such cluster; it might follow from a complete absence of the fragments corresponding to non-coding areas of the genome.

\section{Conclusion}
In conclusion, we briefly list the questions that still await the answers, concerning the transcriptome statistical properties of Siberian larch.
\begin{list}{\arabic{n}.}{\usecounter{n}\leftmargin=9mm \labelwidth=5mm \topsep=0mm \labelsep=2mm \itemsep=1pt \parsep=0mm \itemindent=-1pt}
\item Here we report the results of $K$-means clustering for $K=2$. The clusters produced with the procedure exhibit very interesting properties manifested in the symmetry of the contigs comprising the clusters. Nonetheless, one may develop the clustering for $K=3$, $K=4$, etc. Suppose, one has developed a series of such clusterings from $K=2$ to, say, $K=10$. So the question is what is a pattern of the distribution of contigs over the clusters, as $K$ increases from 2 to~10? Is there any order and a kind of inheritance in the cluster composition, or not? Same question takes place for the contigs of the main genome.
\item What is the reason(s) for transcriptome contigs to exhibit dramatically opposite pattern in the second Chargaff's parity rule execution, in comparison to the contigs of the main genome?
\item What is the reason of octahedral structure appearance for transcriptome?
\item Consider again Fig.~\ref{fig7}; we definitely know that the pattern consists of two oppositional triangles so that these latter gather the contigs from opposite strands. Nevertheless, what is a ``fine'' structure of those triangles? What kind of genes or other coding sites occupy the vortexes?
\end{list}
\begin{acknowledgments}
This study was supported by a research grant No~14.Y26.31.0004 from the Government of the Russian Federation.
\end{acknowledgments}

\end{document}